\documentclass[12pt]{article}
\usepackage{lineno}
\usepackage{epsfig}
\usepackage{graphicx}
\usepackage{amsmath}
\usepackage{hhline}
\usepackage{amssymb}
\usepackage{times}
\usepackage{cite}

\newlength{\dinwidth}
\newlength{\dinmargin}
\begin{document}  
\newcommand{\pom}{{I\!\!P}}
\newcommand{\reg}{{I\!\!R}}
\newcommand{\slowpi}{\pi_{\mathit{slow}}}
\newcommand{\fiidiii}{F_2^{D(3)}}
\newcommand{\fiidiiiarg}{\fiidiii\,(\beta,\,Q^2,\,x)}
\newcommand{\n}{1.19\pm 0.06 (stat.) \pm0.07 (syst.)}
\newcommand{\nz}{1.30\pm 0.08 (stat.)^{+0.08}_{-0.14} (syst.)}
\newcommand{\fiidiiiful}{F_2^{D(4)}\,(\beta,\,Q^2,\,x,\,t)}
\newcommand{\fiipom}{\tilde F_2^D}
\newcommand{\ALPHA}{1.10\pm0.03 (stat.) \pm0.04 (syst.)}
\newcommand{\ALPHAZ}{1.15\pm0.04 (stat.)^{+0.04}_{-0.07} (syst.)}
\newcommand{\fiipomarg}{\fiipom\,(\beta,\,Q^2)}
\newcommand{\pomflux}{f_{\pom / p}}
\newcommand{\nxpom}{1.19\pm 0.06 (stat.) \pm0.07 (syst.)}
\newcommand {\gapprox}
   {\raisebox{-0.7ex}{$\stackrel {\textstyle>}{\sim}$}}
\newcommand {\lapprox}
   {\raisebox{-0.7ex}{$\stackrel {\textstyle<}{\sim}$}}
\def\gsim{\,\lower.25ex\hbox{$\scriptstyle\sim$}\kern-1.30ex%
\raise 0.55ex\hbox{$\scriptstyle >$}\,}
\def\lsim{\,\lower.25ex\hbox{$\scriptstyle\sim$}\kern-1.30ex%
\raise 0.55ex\hbox{$\scriptstyle <$}\,}
\newcommand{\pomfluxarg}{f_{\pom / p}\,(x_\pom)}
\newcommand{\dsf}{\mbox{$F_2^{D(3)}$}}
\newcommand{\dsfva}{\mbox{$F_2^{D(3)}(\beta,Q^2,x_{I\!\!P})$}}
\newcommand{\dsfvb}{\mbox{$F_2^{D(3)}(\beta,Q^2,x)$}}
\newcommand{\dsfpom}{$F_2^{I\!\!P}$}
\newcommand{\gap}{\stackrel{>}{\sim}}
\newcommand{\lap}{\stackrel{<}{\sim}}
\newcommand{\fem}{$F_2^{em}$}
\newcommand{\tsnmp}{$\tilde{\sigma}_{NC}(e^{\mp})$}
\newcommand{\tsnm}{$\tilde{\sigma}_{NC}(e^-)$}
\newcommand{\tsnp}{$\tilde{\sigma}_{NC}(e^+)$}
\newcommand{\st}{$\star$}
\newcommand{\sst}{$\star \star$}
\newcommand{\ssst}{$\star \star \star$}
\newcommand{\sssst}{$\star \star \star \star$}
\newcommand{\tw}{\theta_W}
\newcommand{\sw}{\sin{\theta_W}}
\newcommand{\cw}{\cos{\theta_W}}
\newcommand{\sww}{\sin^2{\theta_W}}
\newcommand{\cww}{\cos^2{\theta_W}}
\newcommand{\trm}{m_{\perp}}
\newcommand{\trp}{p_{\perp}}
\newcommand{\trmm}{m_{\perp}^2}
\newcommand{\trpp}{p_{\perp}^2}
\newcommand{\alp}{\alpha_s}

\newcommand{\alps}{\alpha_s}
\newcommand{\sqrts}{$\sqrt{s}$}
\newcommand{\LO}{$O(\alpha_s^0)$}
\newcommand{\Oa}{$O(\alpha_s)$}
\newcommand{\Oaa}{$O(\alpha_s^2)$}
\newcommand{\PT}{p_{\perp}}
\newcommand{\JPSI}{J/\psi}
\newcommand{\sh}{\hat{s}}
\newcommand{\uh}{\hat{u}}
\newcommand{\MP}{m_{J/\psi}}
\newcommand{\PO}{I\!\!P}
\newcommand{\xbj}{x}
\newcommand{\xpom}{x_{\PO}}
\newcommand{\ttbs}{\char'134}
\newcommand{\xpomlo}{3\times10^{-4}}  
\newcommand{\xpomup}{0.05}  
\newcommand{\dgr}{^\circ}
\newcommand{\pbarnt}{\,\mbox{{\rm pb$^{-1}$}}}
\newcommand{\gev}{\,\mbox{GeV}}
\newcommand{\WBoson}{\mbox{$W$}}
\newcommand{\fbarn}{\,\mbox{{\rm fb}}}
\newcommand{\fbarnt}{\,\mbox{{\rm fb$^{-1}$}}}
%
%
\newcommand{\qsq}{\ensuremath{Q^2} }
\newcommand{\gevsq}{\ensuremath{\mathrm{GeV}^2} }
\newcommand{\et}{\ensuremath{E_t^*} }
\newcommand{\rap}{\ensuremath{\eta^*} }
\newcommand{\gp}{\ensuremath{\gamma^*}p }
\newcommand{\dsiget}{\ensuremath{{\rm d}\sigma_{ep}/{\rm d}E_t^*} }
\newcommand{\dsigrap}{\ensuremath{{\rm d}\sigma_{ep}/{\rm d}\eta^*} }
\newcommand{\dedx}{\ensuremath{{\rm d} E/{\rm d} x}}
\def\Journal#1#2#3#4{{#1} {\bf #2} (#3) #4}
\def\NCA{Nuovo Cimento}
\def\RPP{Rep. Prog. Phys.}
\def\ARNPS{Ann. Rev. Nucl. Part. Sci.}
\def\NIM{Nucl. Instrum. Methods}
\def\NIMA{{Nucl. Instrum. Methods} {\bf A}}
\def\NPB{{Nucl. Phys.}   {\bf B}}
\def\NPPS{Nucl. Phys. Proc. Suppl.} 
\def\NPPSC{{Nucl. Phys. Proc. Suppl.} {\bf C}}
\def\PR{Phys. Rev.}
\def\PLB{{Phys. Lett.}   {\bf B}}
\def\PRL{Phys. Rev. Lett.}
\def\PRD{{Phys. Rev.}    {\bf D}}
\def\PRC{{Phys. Rev.}    {\bf C}}
\def\ZPC{{Z. Phys.}      {\bf C}}
\def\EJC{{Eur. Phys. J.} {\bf C}}
\def\EPL{{Eur. Phys. Lett.} {\bf}}
\def\CPC{Comp. Phys. Commun.}
\def\NP{{Nucl. Phys.}}
\def\JPG{{J. Phys.} {\bf G}} 
\def\EPC{{Eur. Phys. J.} {\bf C}}
\def\PRSL{{Proc. Roy. Soc.}} {\bf}
\def\PETF{{Pi'sma. Eksp. Teor. Fiz.}} {\bf}
\def\JETPL{{JETP Lett}}{\bf}
\def\IJTP{Int. J. Theor. Phys.}
\def\HJ{Hadronic J.}

 


\begin{flushleft}
{\tt \today } \\
\end{flushleft}
\begin{center}
\begin{Large}
{\boldmath \bf Ionization in the atmosphere, comparison between measurements 
and simulations} \\

\end{Large}
 
\begin{flushleft}
 
T.~Sloan$^{1}$, 
G.A. Bazilevskaya$^{2}$,
V.S. Makhmutov$^{2}$,
Y.I. Stozhkov$^{2}$,
A.K. Svirzhevskaya$^{2}$, 
and N.S. Svirzhevsky$^{2}$.          

\bigskip{\it

$^1$ Dept of Physics, University of Lancaster, UK.  \\

$^2$ Lebedev Physical Institute, Moscow, Russia.  \\}

\smallskip

\end{flushleft}
\end{center}


\begin{abstract}
\noindent

A survey of the data on measured particle fluxes and the rate of 
ionization in the atmosphere is presented. Measurements as a function 
of altitude, time and cut-off rigidity are compared with simulations 
of particle production from cosmic rays. The simulations generally 
give a reasonable representation of the data. However, some 
discrepancies are found. The solar modulation of the particle 
fluxes is measured 
and found to be a factor 2.7$\pm$0.8 greater than that observed 
for muons alone near sea level. 

\end{abstract}


 

\section{Introduction}

Ionization in the atmosphere is mainly produced by cosmic rays with a 
component occurring from radioactive elements in the soil.  
The latter dominates the ionization over land at 
altitudes close to sea level. The radiation doses to personnel from 
ionization are usually computed from simulations  
since measurements are not available at all times, altitudes and locations 
on the Earth.  To assess the accuracy 
of the simulations we report in this paper comparisons of the measurements 
of cosmic ray fluxes and ionization rates with the results of the simulations.

\section{The Measurements and Simulations}
A long time series of measurements of particle fluxes in the atmosphere 
at different altitudes has been undertaken by the Lebedev Physical 
Institute (LPI) (Stozhkov et al. 2009). These span the time from 1957 
to the present in the regions of Moscow, Murmansk and Mirny (Antarctica). 
There are also measurements at several other locations 
on the globe for shorter time spans.  The SPARMO Collaboration 
has presented measurements of particle fluxes at different altitudes at 
four locations on the Earth in 1964 (Feiter 1972). Measurements of the 
ionization rates from ion chambers have been presented by by Neher (Neher 1967) 
and by Lowder et al. (Lowder et al. 1972).  

The simulations used are from O'Brien (O'Brien 2005), 
from Usoskin-Kovaltsov (U-K) (Usoskin and Kovaltsov 2006),  
from Berne (Desorgher et al. 2005) and from LPI (Bazilevskaya et al. 2009). 
The U-K simulations are based  on the CORSIKA package and have been extended 
for the upper atmosphere (Usoskin et al 2010). The older U-K version has been 
used here.  The simulations from Berne and LPI are based on the GEANT 4 package.
Both the GEANT 4 and CORSIKA simulations use  
Monte Carlo techniques. The O'Brien simulations are analytic, based on 
solutions of the diffusion equation. 

Comparison of the different measurements at similar cut-off rigidity  
and similar times shows compatibility mostly within 10\% accuracy between 
the LPI and the SPARMO data rising to 20\% at one location.   Comparison 
of the simulations shows compatibility to within 10\% over most 
of the range rising to a 20\% discrepancy between the U-K and O'Brien 
simulations at the highest altitudes (atmospheric depths $<$ 50 g/cm$^2$).

\section{Flux and Ionization}

Some experiments measure omnidirectional particle fluxes whereas others 
measure total ionization rates. The LPI and SPARMO measured the former whilst  
Neher and Lowder et al. measured the latter (Lowder et al. also included 
some flux measurements). The O'Brien simulations give both whilst  
the U-K simulation only gives total ionization. The omnidirectional 
flux, $J$ particles per cm$^2$   
per second, and total ionization, $Q$ ion pairs per cm$^3$ per second,
are related by 
\begin{equation} 
Q=\frac{J <dE/dx>}{\alpha}
\end{equation}  
where $<dE/dx>$ is the average stopping power of all the secondary particles 
produced by the cosmic ray primary and $\alpha$=35eV is the mean energy to 
produce each ion pair (Porter et al. 1976).

\begin{figure}[t]
\vspace*{-10mm}
\hspace*{-10mm}
\begin{center}
\includegraphics[width=10.3cm]{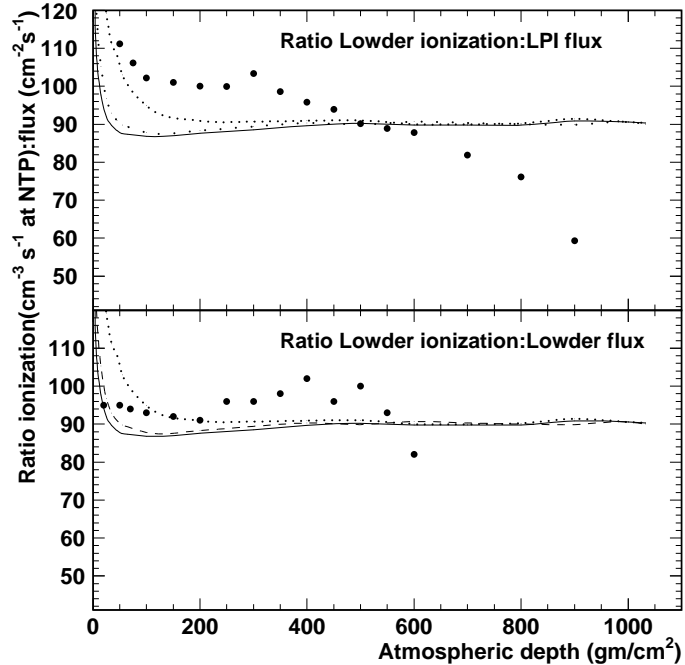}
\end{center}
\vspace*{-10mm}
\label{AAA}
\caption{The ratio of flux to ionization as a function of altitude. 
The points show the measurements.  
The upper plot shows the ratio of the ionization rate measured by Lowder 
et al. to the flux measured by the LPI experiment. These are for the same 
months in 1969-1970 and are interpolated to the same cut-off rigidity. 
The lower plot shows the ratio of the Lowder et al. ionization to the 
Lowder et al. flux (table 3, Lowder et al. 1972). The solid, dashed and 
dotted curves show the ratios expected from the O'Brien simulation at 
solar minimum at cut-off rigidities of R$_C$=2.4, 0.6 and 0 GV, 
respectively.}
\end{figure}
  
Figure 1 shows the measured ratio of $Q/J$ as a function of altitude. 
The ratio is measured to be approximately constant at atmospheric 
depths of less than 600 g/cm$^2$ but falls  
at depths above this. We return to this point 
later. The simulations indicate that the ratio should be constant, but 
with a rapid increase at very high altitude.  Such an effect was 
observed and reported in Stozhkov et al. 2009. If all the particles in the 
shower ionized at the rate of 2 MeV per gm cm$^{-2}$, the minimum of the 
ionization curve,  the ratio should be constant at 74 cm$^{-1}$. 
This is somewhat smaller than the mean value observed for depths less than 
600 g/cm$^2$.    

\section{Comparison of measurements and simulations}

\subsection{Altitude Dependence}

Figure 2 shows a comparison of the particle flux measured by the LPI 
group as a function of altitude averaged over the year 1976. 
The data are compared to the LPI GEANT 4  simulation for the same year 
(a solar minimum). Figure 3 shows the ratios of the measured to the 
simulated values.  There is reasonable agreement between the 
measured data and the simulation except at the highest and lowest 
altitudes (lowest and highest atmospheric depths). 
All the simulations 
show similar discrepancies. The discrepancy at lower depth (high altitude) 
is surprising since the flux here is mainly governed by the primary 
particles. All the simulations model this in a similar way   
using the force field equation  (Gleeson and Axford 1968) which 
has been shown to be a good approximation (Caballero-Lopez and Moraal 2004). 

The discrepancy at highest depth could be due to radioactivity from 
ground based sources. Indeed the historic Hess measurements (Hess 1912) 
show a similar deviation from the simulations at low altitude. 
However, the same discrepancy appears in the data from Mirny in Antarctica.  
This is snow covered all year round so that the contribution from 
Earth sourced radioactivity should be small at that location.   

\begin{figure}[t]
\vspace*{-10mm}
\hspace*{-10mm}
\begin{center}
\includegraphics[width=10.3cm]{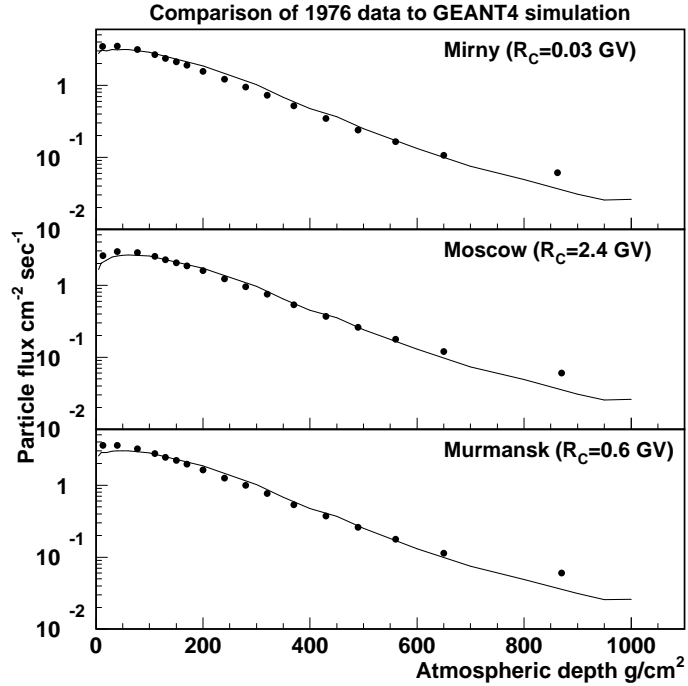}
\end{center}
\vspace*{-10mm}
\label{BBB}
\caption{The LPI data at three different cut-off rigidities (R$_C$) 
showing the particle flux in counts per cm$^2$ per 
second as a function of atmospheric depth in g/cm$^2$ (solid points) 
compared with the LPI simulation (smooth curves).}
\end{figure}

\begin{figure}[htb]
\vspace*{-10mm}
\hspace*{-10mm}
\begin{center}
\includegraphics[width=10.3cm]{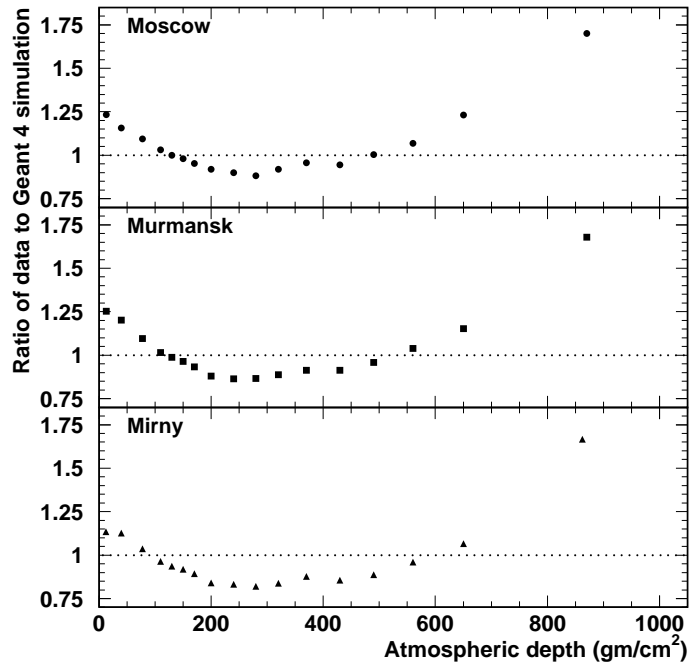}
\end{center}
\vspace*{-10mm}
\label{BBB1}
\caption{The ratios of the LPI measurements of particle flux  
at the cut-off rigidities (R$_C$) shown in figure 2 to the values from 
the LPI simulation.}
\end{figure}

The discrepancy between the LPI data and the simulations at large 
atmospheric depth is not apparent if comparison is made with the ionization 
data from Lowder et al. (O'Brien 2005). However, the latter measurements 
were made with a high pressure ionization chamber. This had a wall thickness 
of 1.1~g/cm$^2$ of steel. Protons of energy less than 30~MeV and electrons 
of energy less than a few MeV cannot penetrate such a wall thickness. 
The LPI data were taken with detectors of wall thickness 0.05~g/cm$^2$ 
of steel with thresholds of $\sim$0.2~MeV for electrons and 5~MeV for protons.
The background noise in the LPI data is less than 10\% of the signal 
at sea level, so this cannot account for the discrepancy.    
The lower energy threshold of the LPI detectors compared to the Lowder 
ion chamber implies that  the discrepancy at low altitude is caused by 
low energy particles. A contribution to the low values of 
the ratio $Q/J$ near sea level, shown in figure 1, could also 
come from such low energy particles.
    
\subsection{Time Dependence}

The time dependence of the particle flux is shown in figure 4. 
The left hand and right hand panels show the measurements and O'Brien 
simulations, respectively, for each month against time. Only selected 
altitudes are shown for clarity. The 11-year solar modulation is 
visible at all altitudes in both the measurements and the simulations 
with an amplitude which increases with altitude.  The discrepancies 
between the absolute values of the fluxes from the simulation and the 
measurements at low and high altitudes are more apparent on the 
linear scale in this plot than on the log scale in figure 2.    

\begin{figure}[t]
\begin{center}
\includegraphics[width=10.3cm]{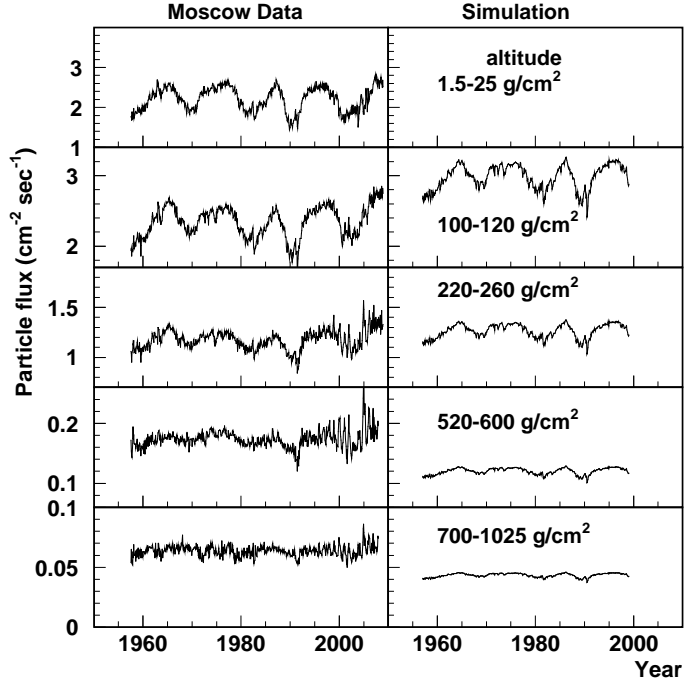}
\end{center}
\vspace*{-10mm}
\label{CCC}
\caption{The particle fluxes in counts per cm$^2$ per 
second from the LPI data as a function of time at various 
atmospheric depths (left hand plots) 
compared to the O'Brien simulations (right hand plots).}
\end{figure}

The times of maximum and minimum count rates were identified in 
figure 4. The times of minimum count rate were 1958.8, 1969.3, 
1982.5, 1990.5 and 2001.3 and those of maximum count rate were 
1965.0, 1976.6, 1986.9, 1996.8 and 2006.5. These dates occur 
slightly later than the corresponding sun spot number peaks  
due to the well known delay in the response of the cosmic rays.     
The data and simulations were then each averaged for 
$\pm$0.5 years on each side of these times. The modulation  
fraction, $f$ is defined by 
\begin{equation}
f = \frac{2 (\rm{Max} - \rm{Min})}{(\rm{Max} + \rm{Min})}
\end{equation}
where Max and Min are the maximum and minimum count rates averaged 
in this way.  The modulation fraction was then computed at each 
altitude in the same way for both the simulation and the measurements.  

Figure 5 shows the resulting modulation fractions, $f$, as a 
function of altitude from the measured data and from the O'Brien 
flux simulation. We also show the modulation of the ionization rate 
from the U-K simulation. It can be seen that the O'Brien simulation 
follows the solar modulation reasonably well whereas the U-K simulation 
predicts a larger modulation than that observed. This discrepancy 
arises as follows.   
The measured LPI data at solar maximum agree well with the flux 
computed from the U-K simulation assuming a constant value of 
$Q/J$=90~cm$^{-1}$. This is a reasonable assumption for depths between 
100-600 g/cm$^2$ (see figure 1). However, at solar minimum the U-K 
simulation predicts a larger flux than that observed, giving too large 
values of $f$.  The discrepancy at solar minimum could be
related to the fact that the modulation potential derived from neutron monitor
data may overestimate the flux of low energy particles below the neutron 
monitor threshold
that may become more noticeable around solar minima (Usoskin 2010a).   

\begin{figure}[t]
\vspace*{2mm}
\begin{center}
\includegraphics[width=10.3cm]{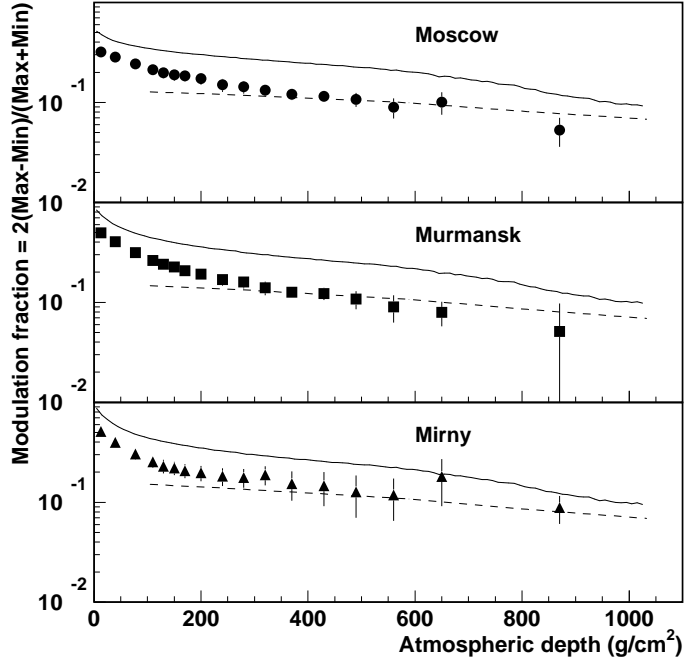}
\end{center}
\vspace*{-10mm}
\caption{The solar modulation fraction from the data and the simulations. 
The solid curve shows the results from the U-K simulation of the ionization 
rate while the dashed 
curve shows those from the O'Brien flux simulations.}
\end{figure}

The measured fractional solar modulation at an altitude of 900 g/cm$^2$ 
averaged over the 3 locations is 
6.5$\pm$1.7 \%.  This agrees well with the value deduced by Sloan and 
Wolfendale~(2008). The value is smaller than the solar modulation fraction 
for neutron monitors at a similar latitude of 15-20\% but larger than that 
for muons. The mean value of this fraction for muons from shielded 
ion chamber data at a similar value of cut-off rigidity was found to 
be 2.4$\pm$0.3\% (Ahluwalia 1997). 
Hence the measured solar modulation for all charged particle fluxes is a 
factor 2.7$\pm$0.8 greater than the value for muons alone. 
The O'Brien simulation predicts that 72\% of the flux of cosmic ray 
particles at this level are muons. Hence the solar modulation of the 
flux from the remaining particles (the soft component of cosmic rays) 
must be larger than that for muons and closer to that seen from neutron 
monitors.      

\subsection{Cut-off rigidity dependence}

The data from the SPARMO collaboration at various places with different 
cut-off rigidity, R$_C$, are shown in figure 6. These data were taken 
at various times in 1964 during solar minimum activity. The solid and 
dashed curves show the predictions of the U-K and O'Brien simulations 
averaged throughout 1964, respectively. To obtain a flux value, the 
U-K ionization simulations have been adjusted assuming a constant $Q/J$ 
ratio of 90 cm$^{-1}$, taken from the data in figure~1. Figure 7 shows 
the ratios of the measured data to the simulations. 

The simulations represent the trend of the data. However, there are 
significant differences with the O'Brien simulations at 
cut-off rigidities above 4.6 GV. 

\begin{figure}[htb]
\vspace*{2mm}
\begin{center}
\includegraphics[width=9.0cm]{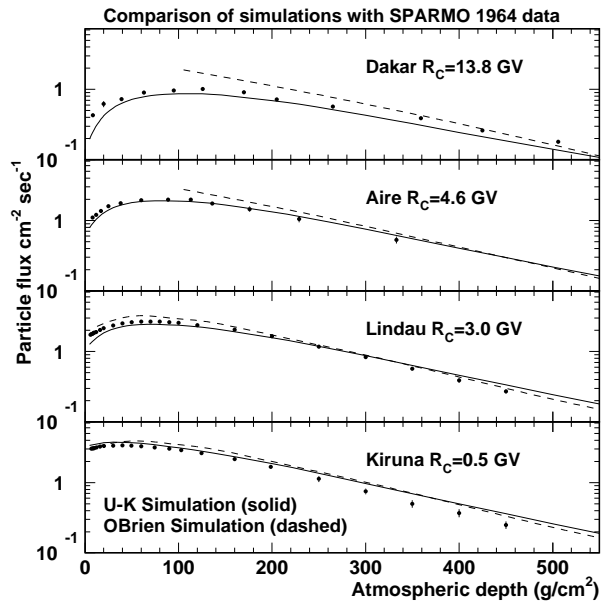}
\end{center}
\vspace*{-10mm}
\caption{The altitude dependence at different cut-off rigidity, R$_C$, 
as measured by the SPARMO collaboration compared with the U-K  
and O'Brien simulations. }
\end{figure}

\begin{figure}[htb]
\vspace*{2mm}
\begin{center}
\includegraphics[width=9.0cm]{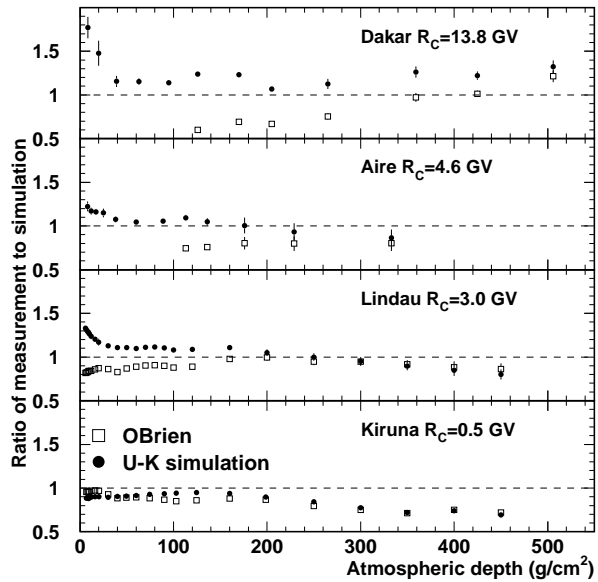}
\end{center}
\vspace*{-10mm}
\caption{The ratios of the SPARMO data at the cut-off 
rigidity, R$_C$, shown in figure 6 to the U-K and O'Brien 
simulations.   }
\end{figure}

\subsection{Long term dependence}

The measured data and the simulations were smoothed over the solar 
cycle using an averaging interval of $\sim$11 years. The interval was 
corrected at different times for the differing solar cycle lengths 
following the procedure described by Lockwood and Fr\"ohlich (2007).  
Figure 8 shows the results. 

\begin{figure}[t]
\vspace*{2mm}
\begin{center}
\includegraphics[width=9.0cm]{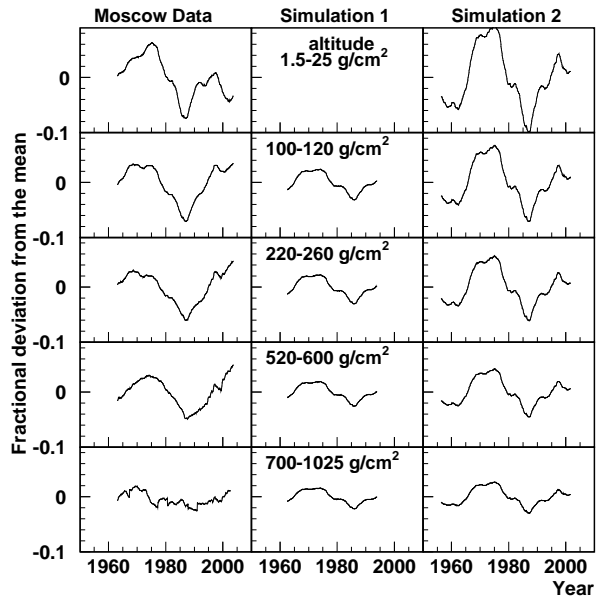}
\end{center}
\vspace*{-10mm}
\caption{Results of 11-year smoothing of the data (left hand plots), 
the O'Brien (simulation 1, centre plots) and the U-K 
simulations (simulation 2, right hand plots) at different altitudes.}
\end{figure} 

The simulations show qualitatively similar behaviour to the measured data 
with a possible 22-year cycle being present in each. The measured data 
show a tendency to increase after the year 2000, perhaps reflecting 
the rather quiet solar behaviour of recent years.

\section{Conclusions}

The simulations and the measured data are in general agreement with 
each other. 
There are discrepancies with the measured data at low altitude where 
radioactivity from ground based sources may be 
expected to contribute. However, the disagreements at this altitude persist 
in Antarctica where the contribution from radioactivity would be expected 
to be low. Perhaps this indicates a contribution from long-lived
atmospheric radioactivity produced by cosmic rays. 
There are also some discrepancies at very high altitude. The 
U-K simulation predicts a solar modulation which is larger than expected 
in the data whereas the O'Brien simulation represents the data well. 
Other discrepancies also occur at cut-off rigidities above 4.6 GV.  

The measurements show that the fractional solar modulation of the 
ionization and the flux is a factor 2.7$\pm$0.8 greater than that observed 
for muons near sea level, obtained from shielded ion chamber measurements. 
This reflects the greater solar modulation of the soft component of the 
cosmic rays than that for the muons (the hard component).

\section{Acknowledgements}

We wish to thank  E. Palle, K. O'Brien and I. Usoskin 
for supplying us with the data from their simulations. We thank  
Arnold Wolfendale for stimulating discussions. TS thanks   
the Dr John Taylor Foundation for financial support. The LPI group 
thanks the Russian Foundation
for Basic Research (grants 08-02-00054, 08-02-91006,
10-02-00326, 10-02-10022k ) and the
Program of Presidium of RAS Neutrino physics and
astrophysics for support.

\end{document}